\documentclass[twocolumn,aps,amssymb,footinbib]{revtex4-1}
\usepackage{amssymb}
\usepackage{graphicx}
\usepackage{amsmath}
\usepackage{times}
\usepackage{color}
\usepackage{subfigure}

\usepackage{bm}

\newcommand{\rr}{\mathbf{r}}
\newcommand{\qq}{\mathbf{q}}

\newcommand{\dd}{\text{d}}
\newcommand{\UU}{\mathcal{U}}
\newcommand{\FF}{\mathcal{F}}
\newcommand{\RR}{\mathbf{R}}
\newcommand{\DDelta}{\bar{\Delta}}
\newcommand{\uu}{\bar{u}}
\newcommand{\Aa}{\tilde{A}}
\newcommand{\da}{\delta \tilde{A}}

\usepackage{times}
\usepackage{amsmath}
\usepackage{amsfonts}
\usepackage{amssymb}
\usepackage{bm}
\usepackage{natbib}
\usepackage{graphicx}
\usepackage{epstopdf}
\usepackage{enumerate}
\usepackage{color}

\graphicspath{{./figs/}{./}}

\begin{document}

\begin{abstract}
\noindent We propose that impurities in Bose-Einstein condensates can serve as a minimal laboratory system to explore the effects of quantum and thermal fluctuations on solvation. Specifically, we show that the role of quantum fluctuations in the formation of solvation shells and the breakdown of linear response theory can be explored in detail.  
\end{abstract}

\title{Bose-Einstein Condensates: a model system for particle solvation?}

\author{Shahriar Shadkhoo$^1$ and Robijn Bruinsma$^{1,2}$}
\affiliation{$^1$Department of Physics and Astronomy, University of California, Los Angeles, CA 90095, USA}\affiliation{$^{2}$Department of Chemistry and Biochemistry, University of California, Los Angeles, CA 90095, USA}

\maketitle
The study of the solvation of particles in fluids is a fascinating area of chemical physics with an enormous literature \cite{Rivail, Marcus} that dates back to two centuries \cite{Grotthuss}. Solvated atoms and ions are surrounded by partially ordered shells of solvent molecules (``solvation shells"). On the other hand, solvent molecules surrounding solvated electrons appear to remain disordered \cite{Hart}. Despite much recent progress, funcamental challenges remain, such as the breakdown of linear-response theory \cite{Schwartz1,Kornyshev}, solvent-specific effects that complicate continuum descriptions \cite{Abel,Cramer} and the role of quantum fluctuations \cite{Washel}. As an example, water | a very important solvent | is a highly correlated liquid characterized by complex, fluctuating patterns of hydrogen-bonding \cite{Franks} which can be viewed as precursors of the freezing transition. The solvation of a particle in water depends on its compatibility with these fluctuations \cite{Gurney}. Path integral simulations of proton solvation in water indicate that the pattern of hydrogen bonding surrounding a solvated proton is subject to strong quantum fluctuations \cite{Marx} while simulations of electrons in water \cite{schwartz2} lead to wave-functions that are suggestive of Anderson localization \cite{Chandler}.

Atomic physics at ultra-low temperatures and cavity quantum electrodynamics are providing novel opportunities for exploring quantum many-body systems. It has been proposed that Bose-Einstein condensates (BECs) hosting foreign atoms could serve as laboratory systems for the study of impurities in quantum many-body systems {\cite{Devreese, Tempere}. BECs support solute particles with the character of a large polaron \cite{Tempere, Vlietinck}. In this article we inquire whether BECs could be used to study the role of \textit{non-specific} thermal and quantum fluctuations effects in the formation of solvation shells and the breakdown of linear response theory. Ultra-cold BECs in transversely pumped cavities exhibit spontaneous self-organization in the form of liquid-crystals and periodic lattices \cite{Nagy1} with a lattice constant that is large compared to atomic length scales (determined instead by the cavity modes \cite{Nagy2}). This transition has been described by a quantum version of the Landau-Brazovskii order-parameter theory (QLB) \cite{Gopal1, Gopal2}, which | in its classical form | has been applied extensively to describe phase transitions between isotropic and periodically modulated phases of block co-polymers and liquid crystals \cite{Brazovskii,Kats,Chaikin}. The general effects of quantum mechanics on solvation should be particularly prominent near a QLB ordering transition, which are known to be subject to strong fluctuations \cite{Gopal1, Gopal2}. We will see that the solvation of a particle in a QLB system near the ordering transition can be addressed analytically and this involves in a natural way the formation of the equivalent of solvation shells and the breakdown of linear response theory.

Let $q_0$ be the preferred wavevector of a periodic ordered phase. In the QLB model, density modes with wavevectors ${\bf{q}}$ that have magnitudes close to $q_0$ are described by a complex order parameter $\rho_{\bf{q}}(t)$ with a Lagrangian
\begin{equation}\label{ff}
\begin{split}
L_{\text{LB}}=&\sum_{\bf{q}}\left(\frac{m_q}{2}\left|\frac{d\rho_{\bf{q}}}{dt}\right|^2-\left[ {(q^2-{q_0}^2)^2+\Delta}\right]|\rho_{\bf{q}}|^2\right)\\&-u\int \dd^3r \left|\rho({\bf r})\right|^4-w\int \dd^3r \left|\rho({\bf r})\right|^6,
\end{split}
\end{equation}
where $m_q$ is the effective mass of the mode. The quadratic mode spectrum has a gap $\Delta$ at $q=q_0$. Next, $u$ measures the strength of the quartic non-linearity and $w$ that of the sixth-order non-linearity \cite{nonlinear}, while $\rho({\bf{r}},t)=\sum_{\bf{q}}\rho_{{\bf{q}}}(t)e^{i{\bf{q.r}}}$. Both $u$ and $w$ are assumed positive while $\Delta$ can have either sign. In mean-field theory, a continuous ordering transition takes place at $\Delta=0$ from a uniform phase to a lamellar liquid crystal phase. Because of the fluctuation effects, the actual phase transition is first-order and takes place at a negative value of $\Delta$ \cite{Gopal1, Gopal2}. 
Next, a solvent or impurity particle of mass $M$ is coupled to the local density of the condensate via a central potential $V(\bf r)$ with a range $a$ that is assumed small compared to $1/q_0$. The Lagrangian of the particle is:
\begin{equation}\label{f-f2}
L_{\text{M}}=\frac{1}{2}M|\dot{\bf R}|^2-\int \dd^3 r\, V({\bf R-r})\left(\rho({{\bf r},t})+{\text{c.c}}.\right).
\end{equation} 
Note that the particle-field interaction locally breaks the +/- symmetry of the QLB Lagrangian. Finally, a lossy environment is included in the Caldeira-Leggett form \cite{Caldeira} as a distribution of harmonic oscillators with a continuous spectrum coupled linearly to the particle:
\begin{equation}\label{f-f2}
L_{\text{E}}=\sum_j\frac{1}{2}m_j\{\dot{x_j}^2-\omega_j^2x_j^2\}-\sum_{j,{\bf{q}}} C_{j,q} {\rho}_{\bf{q}}x_j.
\end{equation}
The nature of the dissipation is determined by the choice of the oscillator spectral density $J_q(\omega)=\frac{\pi}{2}\sum_j\left(\frac{{C_{j,q}}^2}{m_j\omega_j}\right)\delta(\omega-\omega_j)$. We will restrict ourselves to the simplest case of ``Ohmic" dissipation with $J(\omega,q)=\eta_q\thinspace\omega$, for low frequencies where $\eta_q$ is an effective friction coefficient (for higher frequencies an ultra-violet frequency cutoff must be introduced). The classical equation of motion for the particle can be obtained by minimizing the total action, which leads to a Langevin Equation with a damping coefficient that diverges as $\eta_{q_0}/\Delta^2$. 

The equilibrium partition function $\mathcal{Z}$ of the complete many-body system is proportional to the functional integral $\int\exp(\mathcal{S}_T)\mathcal{D}\{{\bf{R}} (t)\}\mathcal{D}\{{\bf{\rho_{\bf{q}}}} (t)\}\mathcal{D}\{{x_j} (t)\}$ over the particle, field, and environmental degrees of freedom. $\mathcal{S}_T$ is the Euclidean action, the integral of the sum of the three Lagrangians over the imaginary ``time" $0<s<\beta$ (with $\beta=1/k_BT$) where $s$ actually has the dimension of inverse energy. All three degrees of freedom must obey periodic boundary conditions in imaginary time, e.g. $\rho_{\bf{q}} (s+\beta)=\rho_{\bf{q}}(s)$. The path integral over the environmental oscillators can be carried out analytically. The remaining path integrals over the density modes and particle trajectories will be discussed separately for positive and negative $\Delta$.\\

{{$\bf{\Delta> 0}$}} .| The effects of the non-linear terms in the field Lagrangian are minor for positive large $\Delta$. If these are dropped then the density fluctuations are harmonic and can be integrated over as well, leading to an effective particle action
\begin{align}
&\mathcal{S}\simeq-\int_0^{\tilde\beta}\frac{1}{2}\left(\frac{\dd{\tilde{\bf{R}}}}{\dd\tilde{s}}\right)^2\dd\tilde{s}-\tilde{{\bf{f}}}.\int_0^{\tilde\beta}\tilde{\bf{R}}(\tilde{s})\dd\tilde{s}\nonumber\\
&+\alpha\int \dd^3\tilde{q}\iint_0^{\tilde\beta}\dd\tilde{s}\dd\tilde{s}'{G^{(2)}_{\tilde{q}}}(|\tilde{s}-\tilde{s}'|)\footnotesize{e^{ i{{\tilde{\bf{q}}}.({\bf{\tilde{R}}}(\tilde{s})-{\bf{\tilde{R}}}(\tilde{s}'))}}}.
\end{align}
We shifted here to dimensionless quantities, indicated by tildes, with mass measured in units of $M$, length measured in units of $1/q_0$, time measured in units of $M/(\hbar q_0^2)$, so energy is measured in units of $(\hbar q_0)^2/M$. In order to later compute the effective mass, an infinitesimal external force $\tilde{\bf{f}}$, is included in the action. The prefactor $\alpha=\frac{M^3|V_{0}|^2}{\pi q_0^3\hbar^4m_{q_0}}$, with $V_{0}=-\int \dd^3r V(r)e^{i\qq_0.\rr}$, is the dimensionless coupling constant. The dimensionless inverse temperature $\tilde\beta=(\hbar q_0)^2/Mk_BT$ is the ratio of the zero-point energy of the particle confined in a well with a dimension of order $1/q_0$ and the thermal energy. Tilde signs will be dropped from hereon. The kernel equals
\begin{align}
G_q^{(2)}(\tau)=\frac{1}{{\beta}}\sum_{n=-\infty}^{+\infty}\frac{e^{i\omega_n \tau}}{{\chi}(q^2-1)^2+{\Gamma}+\gamma |\omega_n|+\omega_n^2}.
\end{align}
The summation is over the dimensionless Matsubara frequencies ${\omega}_n=2\pi n/{\beta}$, so the periodic boundary conditions in imaginary time are obeyed. The dimensionless distance to the mean-field critical point of the QLB is defined here as $\Gamma=\Delta\frac{2M^2}{m_{q_0}\hbar^2 q_0^4}$, the dimensionless friction coefficient as $\gamma=\eta_{q_0}\frac{2M}{\hbar{q_0}^2m_{q_0}}$ and the dimensionless field rigidity as ${\chi}=\frac{2M^2}{\hbar^2 m_{q_0}}$. This action, in effect, corresponds to linear-response theory.

In the limit of zero dissipation $\gamma=0$ | when quantum fluctuations of the field are maximal | the kernel $G_q^{(2)}(\tau)$ decays with $\tau$ as $\exp(-\sqrt{\Gamma}\tau)$ \cite{ft1}. The path integral over the particle degree of freedom then reduces to one that was investigated by Feynman \cite{Feyn} for the Fr{\"o}hlich Hamiltonian, which describes charge carriers in polar crystals coupled to an optical mode. The effective mass of this polaron increases smoothly as a function of $\alpha$. In the opposite limit $\gamma\rightarrow \infty$, when quantum fluctuations of the field are suppressed, only the $n=0$ term remains in the sum of the kernel, which is then independent of $\tau$ and reduces to the classical mean-field static structure factor $\frac{1}{{\chi}(q^2-1)^2+{\Gamma}}$ \cite{ft2}. In Supplementary Material 1 it is shown that the path integral over the particle trajectories reduces to that of a particle in a self-consistent attractive radial potential. If the coupling constant exceeds a threshold, then a bound state appears. This \textit{self-trapping}, or ``small polaron",  concept was first proposed in 1933 by Landau \cite{Landau} again for the case of charge carriers in polar crystals. The concept was developed further by Holstein \cite{Emin} and others, and many examples of self-trapping have been documented in the experimental literature \cite{Stoneham}. 

For general $\gamma$, the path integral over the particle trajectories can be performed variationally. Define a Gaussian trial action:
\begin{align}
\mathcal S_t=& {-\int_0^{\beta}\frac{1}{2}\left(\frac{\dd{{\bf{R}}}}{\dd{s}}\right)^2\dd{s}}-{\bf{f}}.\int_0^{\beta}{\bf{R}}(s)\dd s\nonumber\\
&-\frac{1}{2}\int\int_0^{\beta}\mathcal K(|s-s'|) |{\bf{R}}(s)-{\bf{R}}(s')|^2 \dd s\dd s',
\end{align}
where the kernel $\mathcal K(\tau)=\frac{1}{{\beta}}\sum_{n=-\infty}^{+\infty}\mathcal K_ne^{i\omega_n \tau}$ with
\begin{equation}
\mathcal K_n=\frac{C}{D+\gamma|\omega_n|+\omega_n^2},
\end{equation}
is similar to the actual kernel. The constants $C$ and $D$ play the role of variational parameters that are determined by applying Feynman's inequality for trial actions: $F \leq F_t+\frac{1}{\beta}\langle \mathcal S-\mathcal{S}_t\rangle$. Here ${F(f)=-k_BT} \ln\mathcal{Z}$ is the free energy of the particle. Expectation values are computed using the Gaussian trial action. For a free particle with mass $M^*$ subject to a force $f$, the second derivative of the free energy with respect to the applied force equals $\partial^2F/\partial f^2 |_{f=0}=\hbar^2\beta^2/12M^*$ (in actual units). Using this as the definition of the effective mass \cite{Saitoh} and applying the trial action gives:
\begin{equation}
M^* = \frac{\beta^2}{24}\bigg[\sum_{n>0}g_n\bigg]^{-1},
\end{equation}
(in reduced units) where $g_n=[\omega_n^2+\frac{2}{\beta}(\mathcal{K}_0-\mathcal{K}_n)]^{-1}$. Figure \ref{fig:1a} shows the effective mass as a function of the coupling constant $\alpha$ and the damping coefficient $\gamma$ for $\beta=100$, $\chi=100$, and $\Gamma=1$.
\begin{figure}
\centering
\begin{subfigure}{\label{fig:1a}}
\centering
\includegraphics[width=0.8\linewidth]{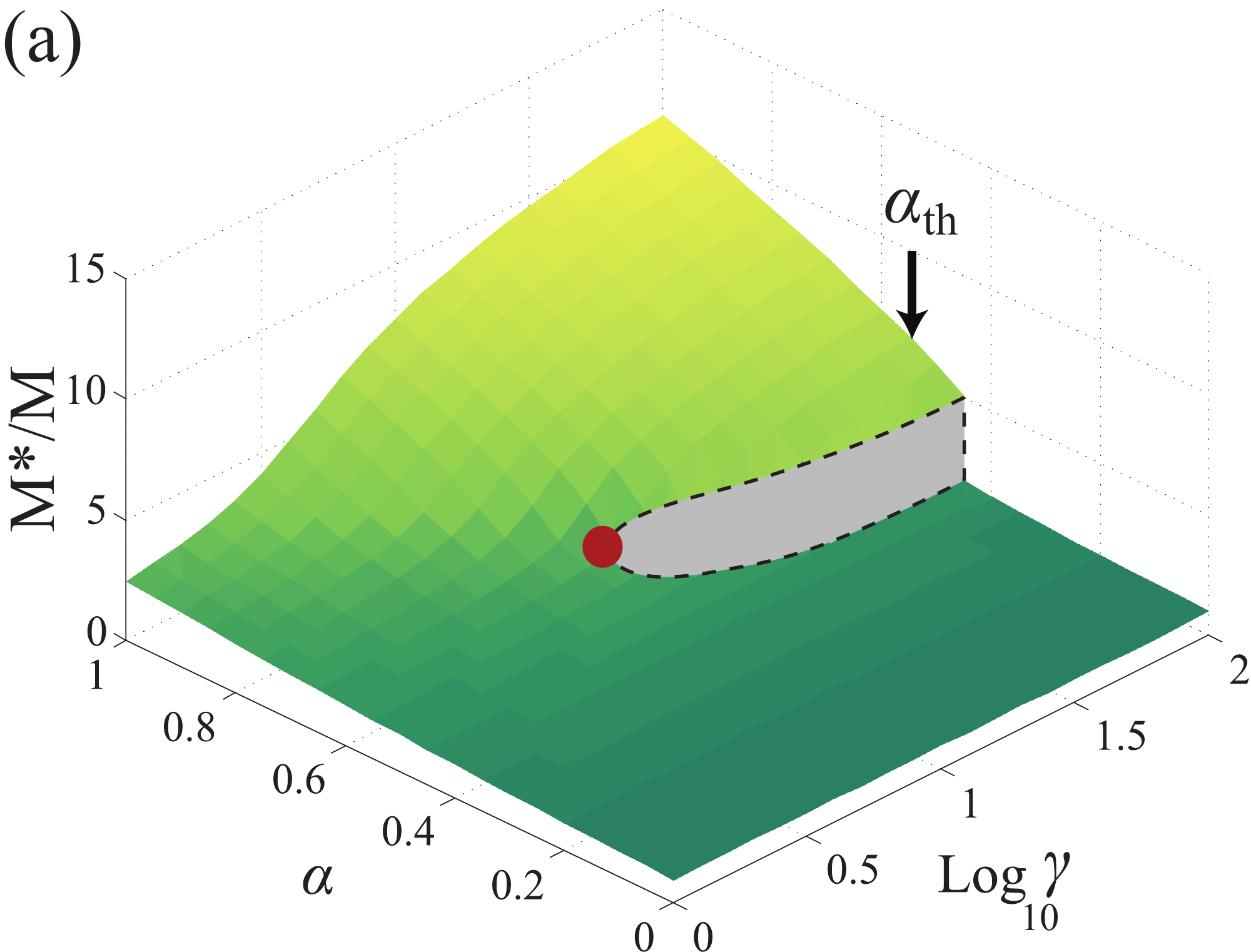}\\
\end{subfigure}
\begin{subfigure}{\label{fig:1b}}
\centering
\includegraphics[width=0.8\linewidth]{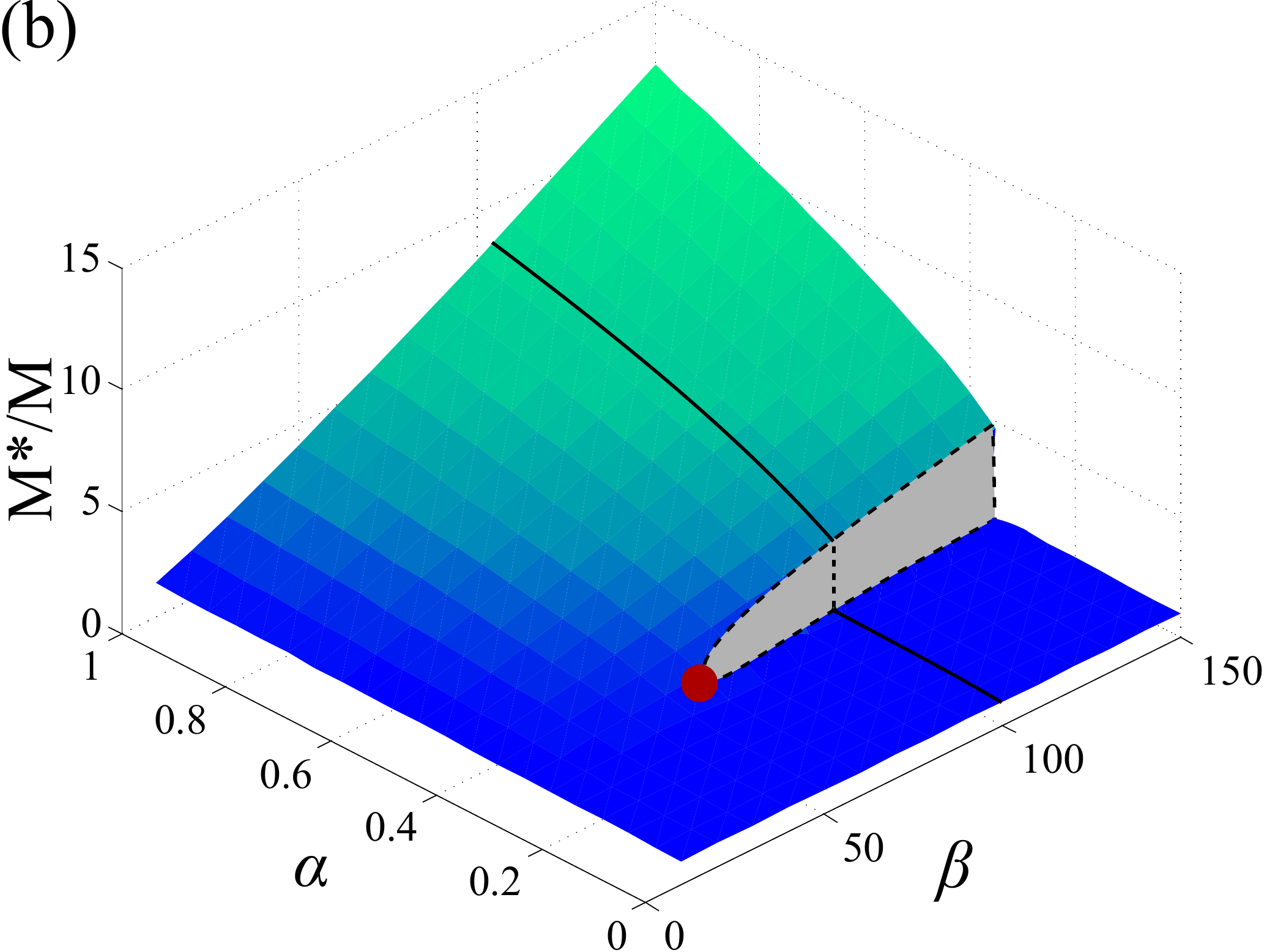}\\
\end{subfigure}
\caption{(a) Effective mass as a function of the dimensionless coupling constant $\alpha$ and the dimensionless friction coefficient $\gamma$ of the ordering field for $\beta=100$ and $\Gamma=1$. The red dot indicates a critical point. The black arrow marks the theoretically predicted $\alpha_{\text{th}}=0.54$ for $\gamma,\beta\to\infty$. (b) Effective mass as a function of $\alpha$ and the inverse temperature $\beta$ for $\Gamma=1$ for large $\gamma$. }
\end{figure}
In the limit of strong dissipation, the effective mass undergoes a first-order discontinuity as a function of the coupling constant $\alpha\simeq0.54$ consistent with Landau self-trapping (Supplementary Material 1). The mass enhancement strongly decreases as the dissipation level is reduced while the limit of weak dissipation, the mass enhancement is in agreement with Feynman's results. The two regimes are separated by a \textit{critical point}, indicated by a red dot \cite{AP}. Figure \ref{fig:1b} shows the temperature dependence of the mass enhancement: thermal fluctuations suppress the self-trapped state. The critical point is now at finite temperature. In  Supplementary Material 2 we show that the free energy $F$ has a slope discontinuity along the same lines in the $\alpha-\gamma$ and $\alpha-\beta$ phase planes where the effective mass has a jump discontinuity and that the free energy exhibits metastability near the line, characteristic of a first-order transition. The effective mass of the self-trapped particle, increases with decreasing temperature approximately as $a_1\beta+a_2\beta^2$ with $a_{1,2}$ constants. This can be understood from the fact that the Lagrangian has continuous translation symmetry so a self-trapped particle can diffuse at finite temperature under the action of the thermal fluctuations of the medium with a diffusion coefficient that vanishes in the limit of low temperature proportional to $1/\beta\gamma$ according to the fluctuation-dissipation theorem. 

\vspace*{1\baselineskip} 

{{$\bf{\Delta\lesssim0}$}} .| Near the mean-field stability limit $\Delta=0$, the role of quantum and thermal fluctuations become so important that the quartic and sixth order non-linearities of the QLB Lagrangian must be taken into account. Now, before performing the functional integral over the field and particle degrees of freedom, we first expand the free energy $F({\{\bf{R}}(s)\})$ for a given particle trajectory ${\bf{R}}(s)$ in a Taylor expansion in function space in powers of the interaction potential $V(r)$ and then perform the functional integral over the field. The zero-order term is then the partition function in the absence of the particle while the first-order term is
\begin{align}
&\Delta F^{(2)}({\{\bf{R}}(s)\}) = \nonumber\\& -\frac{1}{2!}\sum_{\qq,n}\mathcal{G}_q^{(2)}(\omega_n) |V_q|^2\iint_0^{\beta}\dd{s}\dd{s}'e^{{i\qq.({\bf{R}}(s)-{\bf{R}}(s'))}},
\end{align}
\noindent where $\mathcal{G}_q^{(2)}(\omega_n)$ is the full two-point Green's function of the system, including fluctuation corrections due to the non-linear terms, but in the absence of the particle. Similarly, the second-order term in the expansion contains the full four-point vertex function of the system, again in the absence of the particle. 

The full two, four-point, and higher-order correlation functions, are obtained from diagrammatic expansion of the impurity-free theory. To one-loop order, this leads in the low-temperature limit to a frequency and momentum independent renormalization $\bar{\Delta}\simeq\Delta+\mathcal P u\ln(\Delta_c/\bar\Delta)$ of the gap in the excitation spectrum and a renormalization $\bar{u}\simeq u\frac{1-u\Pi}{1+u\Pi}$ of the coefficient of the quartic term, with $\mathcal P\propto {q_0}^2$, $\Delta_c$ a high-energy cutoff, and $\Pi=\mathcal P/{\bar{\Delta}}$  \cite{Gopal2}. The renormalized gap $\bar\Delta$ of the excitation spectrum decreases for negative $\Delta$ but remains positive, while the renormalized quartic coefficient $\bar u$ becomes negative when $\bar\Delta$ drops below $\mathcal P u$. 

A functional integral over the field degrees of freedom for the case of an action with this renormalized Lagrangian | including the impurity | no longer suffers from strong fluctuations since $\bar\Delta$ is now positive. An expansion of this functional integral in powers of the impurity potential, but now neglecting fluctuation corrections, reproduces the earlier expansion so one can effectively adopt the renormalized Lagrangian and neglect fluctuation corrections. The functional integral over the field can thus be performed by expanding around the stationary points of the action, which correspond to solutions of the classical equation of motion for the field. 

First consider the case with the particle fixed at the origin, a static impurity. The classical equation of motion for the order-parameter field then has a time-independent solution in the form of a radial density modulation $\rho(r)=A(r)e^{i\qq_0.\rr}$. Away from the origin, the modulation amplitude $A(r)$ is the solution of the second order non-linear differential equation
\begin{equation}\label{classical}
\frac{\dd^2A}{\dd r^2}+\frac{2}{r}\frac{\dd A}{\dd r}\simeq\frac{1}{q_0^{2}} \,\frac{\dd}{\dd A}\left[\bar{\Delta}|A|^2+\bar{u}|A|^4+w|A|^6)\right],
\end{equation}
which must be solved under the condition that $A(r)$ goes to zero far from the origin. The amplitude $A(r=0)$ at the origin, where the field interacts with the particle, is kept as a free parameter. In Supplementary Material 3 it is shown that, close to the ordering transition, the energy of a radial profile obeys the scaling form $U(A^*_{\text{s}})=\frac{{\xi^3\bar{\Delta}}^2}{|\bar{u}|} g(A^*_{\text{s}})$ where $\xi\equiv2q_0/\sqrt{\bar{\Delta}}$ is the correlation length at the ordering transition and where $A^*_{\text{s}}=(\bar{u}/\bar{\Delta})^{1/2}A(r=0)$ is the dimensionless modulation at the origin. The dimensionless scaling function $g(x)$, which increases monotonically with $x$, resembles a double-well potential that has been tilted counterclockwise (Supplementary Material 3.2). If the modulation amplitude $A^*_{\text{s}}$ at the origin is treated as a collective coordinate representing the order-parameter field then it has an effective Lagrangian:
\begin{equation}\label{classical2}
L_A=\frac{1}{2}\frac{m_{q_0}\xi^3\bar{\Delta}}{|\bar{u}|}\left(\frac{\dd A^*_{\text{s}}}{\dd t}\right)^2-U(A^*_{\text{s}}),
\end{equation}
while the particle Lagrangian is now
\begin{equation}\label{f-f2}
L_{\text{M}}=\frac{1}{2}M|\dot{\bf R}|^2+(\bar{\Delta}/\bar{u})^{1/2}A^*_{\text{s}}V_0\cos(q_{0}R).
\end{equation} 
The modulation amplitude is, as before, also coupled to the dissipative background. In the strong dissipation limit, integrating over the particle degree of freedom for fixed $A^*_{\text{s}}(t)$ leads to the result that one can replace $U(A^*_{\text{s}})$ by
\begin{align}\label{f-f2}
U_{\text{eff}}(A^*_{\text{s}}) \simeq & \thinspace U(A^*_{\text{s}})-(\bar{\Delta}/\bar{u})^{1/2}V_0A^*_{\text{s}}\nonumber\\
&+ \frac{3}{2}\hbar q_0\sqrt{\frac{(\bar{\Delta}/\bar{u})^{1/2}V_0A^*_{\text{s}}}{M}},
\end{align} 
for $(\bar{\Delta}/\bar{u})^{1/2}V_0{A^*_{\text{s}}}>(9/4)\hbar^2 {q_0}^2/M$. If the inequality does not hold, then $U_{\text{eff}}(A^*_{\text{s}})=U(A^*_{\text{s}})$. The two cases correspond to the presence, respectively, absence of a bound state of the particle in the potential well generated by the modulation (see Supplementary Material 1). In the limit of small $V_0$, where there is no bound state, $U_{\text{eff}}(A^*_{\text{s}})=U(A^*_{\text{s}})$ has a single minimum at $A^*_{\text{s}}=0$. The physical properties of the particle are in this regime the same as those of the large polaron of the previous section with all bare parameters replaced by their renormalized values. Upon increasing $V_0$, a second minimum appears discontinuously with $A^*_{\text{s}}$ proportional to $V_0$. The critical value of $V_0$ is
\begin{equation}\label{h2s}
V_{c1}=\frac{3}{2}\sqrt{\frac{27}{4}\frac{\hbar^2 {q_0}^2}{M} \xi^3\bar{\Delta}}\;.
\end{equation} 
This second state corresponds to the self-trapped state of the previous section. With increasing $V_0$ a third minimum appears as well, now at $A^*_{\text{s}}=\sqrt{2}$. The critical value of $V_0$ equals 
\begin{equation}\label{h2s}
V_{c2}=\frac{9}{4\sqrt{2}}\frac{\hbar^2 {q_0}^2}{M}(\bar{u}/\bar{\Delta})^{1/2}.
\end{equation} 
The new minimum corresponds to a \textit{soliton} solution with $A(r)$ interpolating between the modulation amplitudes of the ordered phase and the uniform phase. Soliton states of a field theory in three dimensions are normally unstable, by Derrick's Theorem, but in Supplementary Material 3.3 it is shown that the impurity potential stabilizes the soliton. The soliton can be viewed as a ``droplet" of modulated material with a size of the order of $\xi$ surrounding the origin, with the particle confined in the center (see Fig. 2).
\begin{figure}[ht]{\label{fig:solvationshell}}
\centering
\includegraphics[width=1\linewidth]{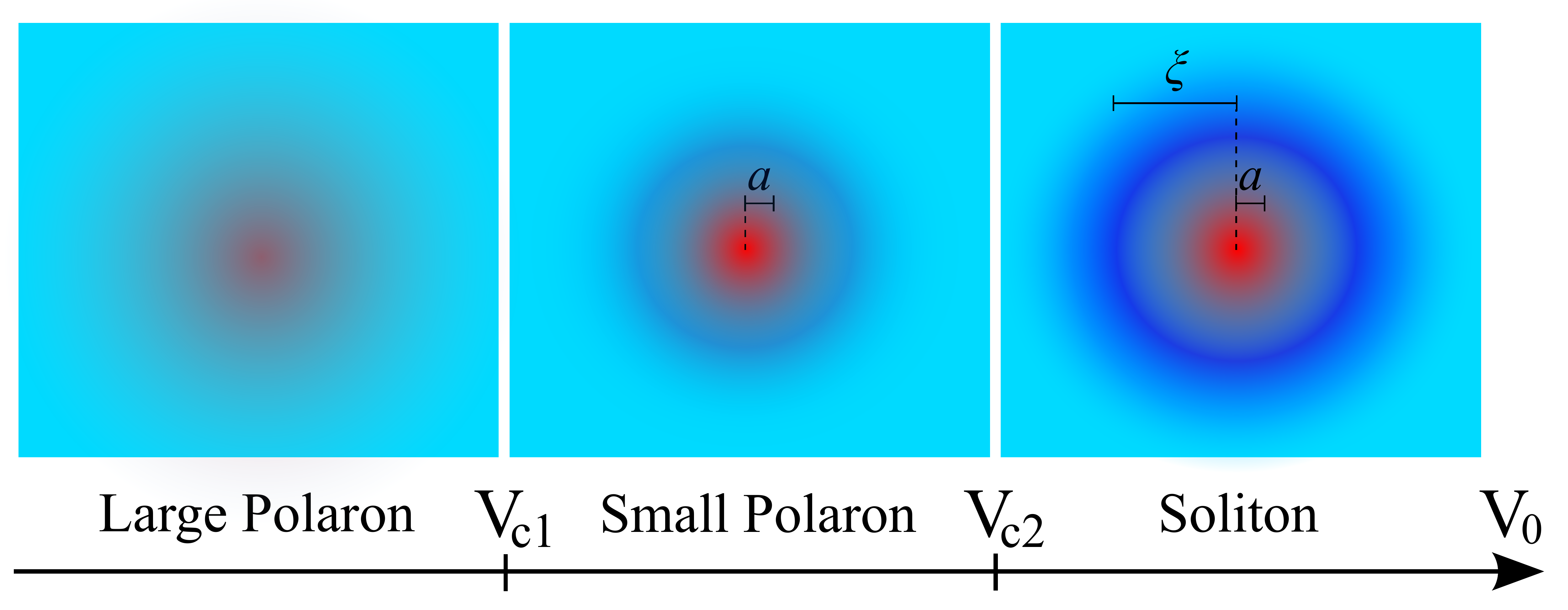}
\caption{\label{fig:solve} The sketch of the ground-state of the BEC-impurity system as a function their coupling $V_0$. The background schematically shows the modulating amplitude $A(r)$, and the red cloud indicates the particle. The left panel shows an extended particle (Large Polaron). The middle panel is a self-trapped particle with a small modulation of the field around it (Small Polaron). The localized (Solvated) particle inside the soliton (solvation shell) is shown in the right panel.}
\end{figure}

Physically, the formation condition of the third minimum means that the work by the particle potential upon formation of the soliton state must exceed the zero point energy of the particle confined inside a potential well with a radius of the order of $1/q_0$. The formation mechanism of the second solution is in principle similar to that of the small polaron except that the soliton state now describes the deformation of the medium. At the formation threshold, the soliton state is metastable. With increasing $|V_0|$, the energy of the soliton state drops below that of the large polaron state at a third threshold $V_{c3}$ where $U_{\text{eff}}(\sqrt{2})\simeq0$ with 
\begin{equation}\label{h3}
V_{c3}- V_{c2}\simeq(\bar{u}/\bar{\Delta})^{1/2}\;\frac{g(\sqrt{2})}{\sqrt{2}}\frac{{\xi^3\bar{\Delta}}^{2}}{{|\bar{u}|}}.
\end{equation} 
Note that the right hand side is the energy scale of $U(A^*_{\text{s}})$. This condition means that the work by the impurity potential during the formation of the soliton must exceed the sum of the zero-point energy and the energy cost of the soliton in the absence of the particle for the soliton state to be the minimum energy state. The weak dissipation limit | in which the dynamics of the non-linear order-parameter field must be explicitly integrated over | is mathematically challenging and will be addressed in a separate publication.

In summary, we find that for the BEC model system, the Landau-Brazovzkii model predicts that impurity particles can adopt a variety of structures. The particle can have the properties of a large polaron or a self-trapped small polaron, depending on the coupling constant, the level of dissipation, and the temperature with a phase diagram that contains a line of first-order transitions ending at a critical point. Close to the ordering transition, where the fluctuations of the order-parameter field become pronounced, a third state appears: the self-trapped soliton state, where the particle is surrounded by a droplet of ordered material. The theory predicts that the solvation state should be strongly dependent on the particle mass. For particles with large mass, the small polaron and soliton states dominate. Both the  $V_{c2}$ and $V_{c1}$ thresholds go to zero. For light particles, the soliton stability threshold $V_{c2}$ diverges as $1/M$ and the large and small polaron states are expected to dominate. In terms of the phenomenology of solvation, the concentric solvation shells of ions in water would be related to the formation of the soliton state. Within the model system then, the formation of solvation shell for heavier particles is mathematically linked to the breakdown of linear-response theory. Lighter particles, like electrons, would be expected to form large or small polarons. A recent mixed quantum-classical simulation of electron solvated in water, that included solvent dynamics, reported a local organization of the water molecules that appears to be quite consistent with Landau self-trapping \cite{schwartz2}. This is encouraging as it suggests that some of our results may extend beyond BEC systems. It should be kept in mind however that the ordered phase of the model system discussed in this letter is a lamellar liquid crystal, not a periodic crystal, but by including a sufficiently strong cubic non-linearity in the Lagrangian, the ordered phase transforms from a lamellar liquid crystal to a true crystal with three dimensional periodicity.  Nevertheless, experimental studies of solute particles in a BEC under conditions where the cubic term is absent should be very informative of the role of fluctuations in solvation phenomena.

Acknowledgments. We would like to thank Sudip Chakravarty, David Chandler, Alexander Grosberg, Eric Hudson, and Ben Schwartz for helpful discussions.

\newpage

\section{Supplementary Materials}

\subsection{Large $\gamma$ Limit and $\Delta>0$}

When quantum fluctuations of the field are suppressed by taking the limit $\gamma\to\infty$, the only Matsubara frequency contributing to the summation in temporal kernel is the $n=0$ term. This corresponds to the Born-Oppenheimer approximation where a quantum particle interacts with a quasi-static configuration of the surrounding medium. In this limit, the kernel is independent of time and reduces to $G_q=\frac{\beta^{-1}}{\chi(q^2-1)^2+\Gamma}$. The effective action for the particle reads:
\begin{align}
\mathcal S=&-\int_0^{\beta}\dd s\,\frac{1}{2}\bigg|\frac{\dd \RR}{\dd s}\bigg|^2\nonumber\\
&+\alpha\int\dd^3q\;G_q\int_0^{\beta}\int_0^{\beta}\dd s\dd s' \exp[i\qq.(\RR(s)-\RR(s'))].
\end{align}
We need a criterion for the appearance of a self-trapped state. Assume that the trajectories $\RR(s)$ that  dominate the path integration in the partition function $\mathcal Z=\int \mathcal D[\RR(s)]\exp(\mathcal S)$ are confined isotropically in a spherical region around the origin and then, \textit{a-posteriori}, verify the assumption. At low temperatures, the factor $f_q\equiv \beta^{-1}\int_0^{\beta}\dd s' \exp[i\qq.\RR(s')]$ samples a long trajectory and thus can then depend only on the magnitude $q$ of the wavevector. The action can be written as:
\begin{align}
\mathcal S=&-\int_0^{\beta}\dd s\,\frac{1}{2}\bigg|\frac{\dd \RR}{\dd s}\bigg|^2\nonumber\\
&+\alpha\beta\int\dd^3q\;G_qf_q\int_0^{\beta}\dd s \exp[-i\qq.\RR(s)].
\end{align}
A self-consistency condition for $f_q$ is then
\begin{align}
\frac{\dd}{\dd\alpha}F(\alpha)=-\beta\int\dd^3q\;G_q|f_q|^2
\end{align}
with $F=-1/\beta \ln\mathcal Z$. After integrating over the angular direction of the wavevector, the action reduces to
\begin{align}
\mathcal S=-\int_0^{\beta}\dd s\,\frac{1}{2}\bigg|\frac{\dd \RR}{\dd s}\bigg|^2 -\int_0^{\beta}\dd s\ U(R(s)).
\end{align}
where
\begin{align}
U(R)=4\pi\alpha\beta\int_0^{\infty}\dd q\,q\,G_qf_q\frac{\sin(qR)}{R}.
\end{align}
This expression has the form of the action of appearing in the path-integral expression of the free energy of a particle moving in the radial potential $U(R)$. If this radial potential has one or more bound states then the lowest bound state is isotropic and the path integral indeed would be dominated at low temperatures by isotropic trajectories, as assumed. In order to determine whether there are bound states, note that the integration over $q$ is dominated (in dimensionless space) by $q=1$ since $G_q$ is peaked at $q=1$ close to the transition. It follows that
\begin{equation}
U(R)\approx\; \frac{2\sqrt{2}\pi^2\alpha f_1}{\sqrt{\chi\Gamma}}\frac{\sin R}{R}.
\end{equation}
For negative $f_1$ this represents a potential well near the origin | the case of interest | and for positive $f_1$ a repulsive potential. For small $R$, one can expand $U(R)$ to the second order in $R$, which leads for negative $f_1$ to a three dimensional harmonic oscillator potential:
\begin{equation}
U(R)\approx\; \frac{2\sqrt{2}\pi^2\alpha f_1}{\sqrt{\chi\Gamma}}(1-R^2/3!+...).
\end{equation}
The ground-state energy level of the harmonic oscillator lies a distance $\Delta\epsilon=\frac{3}{2}\hbar\omega_0$ above $U(0)$, where $\omega_0=\sqrt{\frac{4\pi\alpha |f_1|}{3\sqrt{\chi\Gamma}}}$. An approximate condition for the existence of at least one bound state is obtained by demanding that the lowest energy level $E(\alpha)=U(0)+\Delta\epsilon$ of the harmonic oscillator is negative. Here
\begin{equation}
 E(\alpha)=-\frac{2\sqrt{2}\pi^2 \alpha |f_1|}{\sqrt{\chi\Gamma}}+\frac{3}{2}\sqrt{\frac{2\sqrt{2}\pi^2 \alpha |f_1|}{3\sqrt{\chi\Gamma}}},
\end{equation}
and consequently,
\begin{equation}{\label{criterion}}
\alpha_c>\frac{3\sqrt{2}}{16\pi^2}\frac{\sqrt{\chi\Gamma}}{|f_1|}.
\end{equation}
In the low temperature limit, with $F(\alpha)\simeq E(\alpha)$, the self-consistency condition for $f_1$ reduces to
\begin{equation}{\label{criterion}}
 -{2\sqrt{2}\pi^2 |f_1|}+\frac{3}{4}\sqrt{\frac{2\sqrt{2}\pi^2 \sqrt{\chi\Gamma}|f_1|}{3\alpha}}=-{2\sqrt{2}\pi^2|f_1|^2}.
\end{equation}
Recalling that the minimum $\alpha$ value for a bound-state is $\alpha=\frac{3\sqrt 2}{16\pi^2}\frac{\sqrt{\chi\Gamma}}{|f_1|}$ and inserting this into the self-consistency condition gives
\begin{equation}{\label{criterion}}
\sqrt{2} |f_1|=2\sqrt{2}|f_1|^2,
\end{equation}
with solutions $f_1=0$ and $f_1=-1/2$. Taking the second solution to be the bound-state gives the final result
\begin{equation}{\label{weak}}
\alpha_c>\frac{3\sqrt{2}}{8\pi^2}\sqrt{\chi\Gamma}
\end{equation}
\\

\subsection{Free energy and effective mass plots}

The free energy of the particle $F$ is plotted versus the coupling constant $\alpha$ and (a) ${\text{Log}}_{10}\gamma$ , (b) $\beta$. 
\begin{figure}[ht]{\label{sup1}}
\centering
\begin{subfigure}{\label{fig:sup1a}}
\centering
\includegraphics[width=0.7\linewidth]{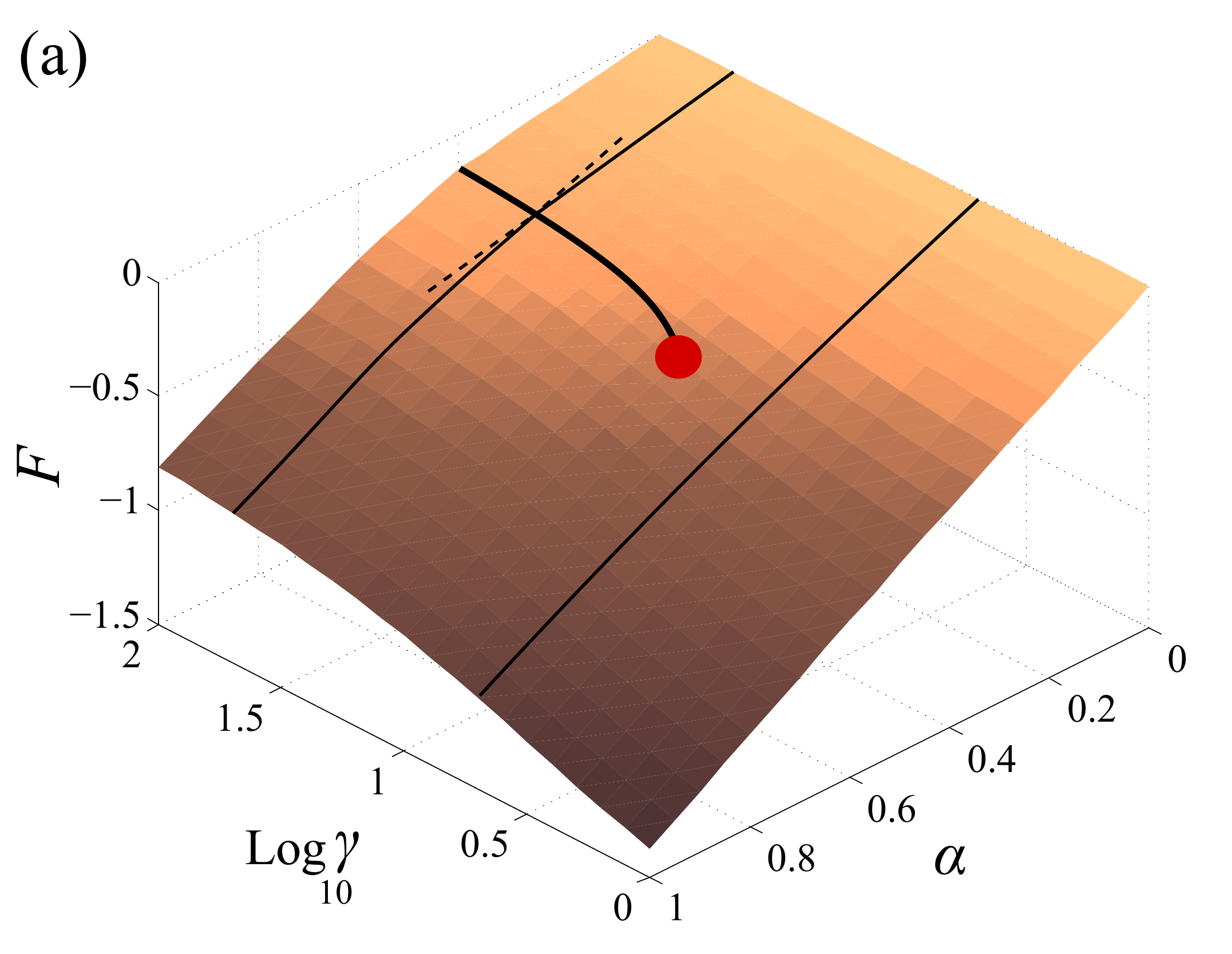}\\
\end{subfigure}
\begin{subfigure}{\label{fig:sup1b}}
\centering
\includegraphics[width=0.7\linewidth]{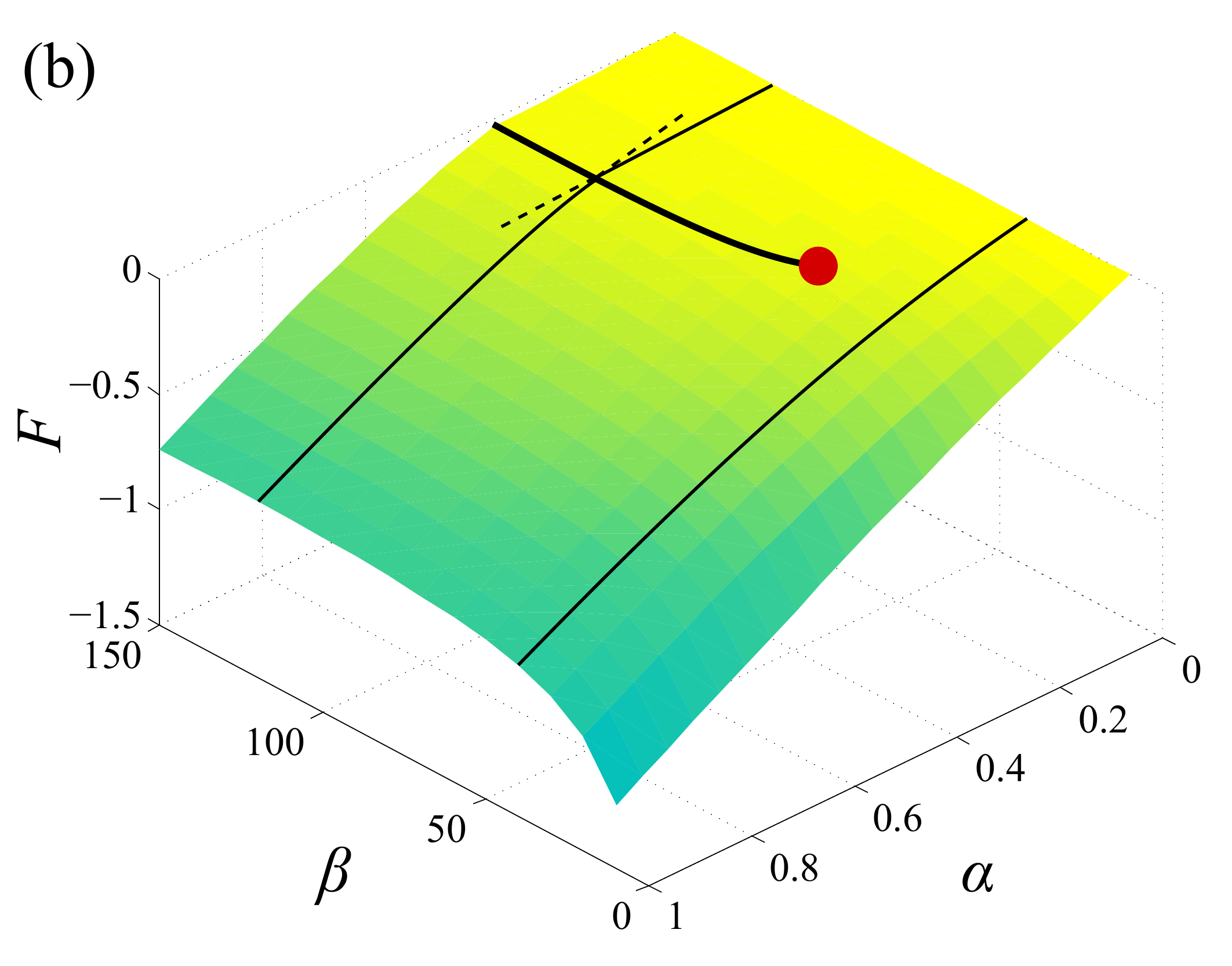}\\
\end{subfigure}
\caption{Minimized free energy plotted in (a) versus $\alpha$, $\gamma$ and in (b) versus $\alpha$ and $\beta$. Along the thick solid line, the derivative of the free energy with respect to $\alpha$ has a discontinuity indicating a first-order phase transition. The line coincides with that of the effective mass discontinuity as shown in the next figure. The red dot at the end of the line marks a critical point. The ends of the dashed segments represent the limits of metastability. }
\end{figure}
The effective mass plotted as a function of the coupling constant and the distance to the critical point is shown in the next figure.
\begin{figure}[ht]{\label{fig:sup2}}
\centering
\includegraphics[width=0.7\linewidth]{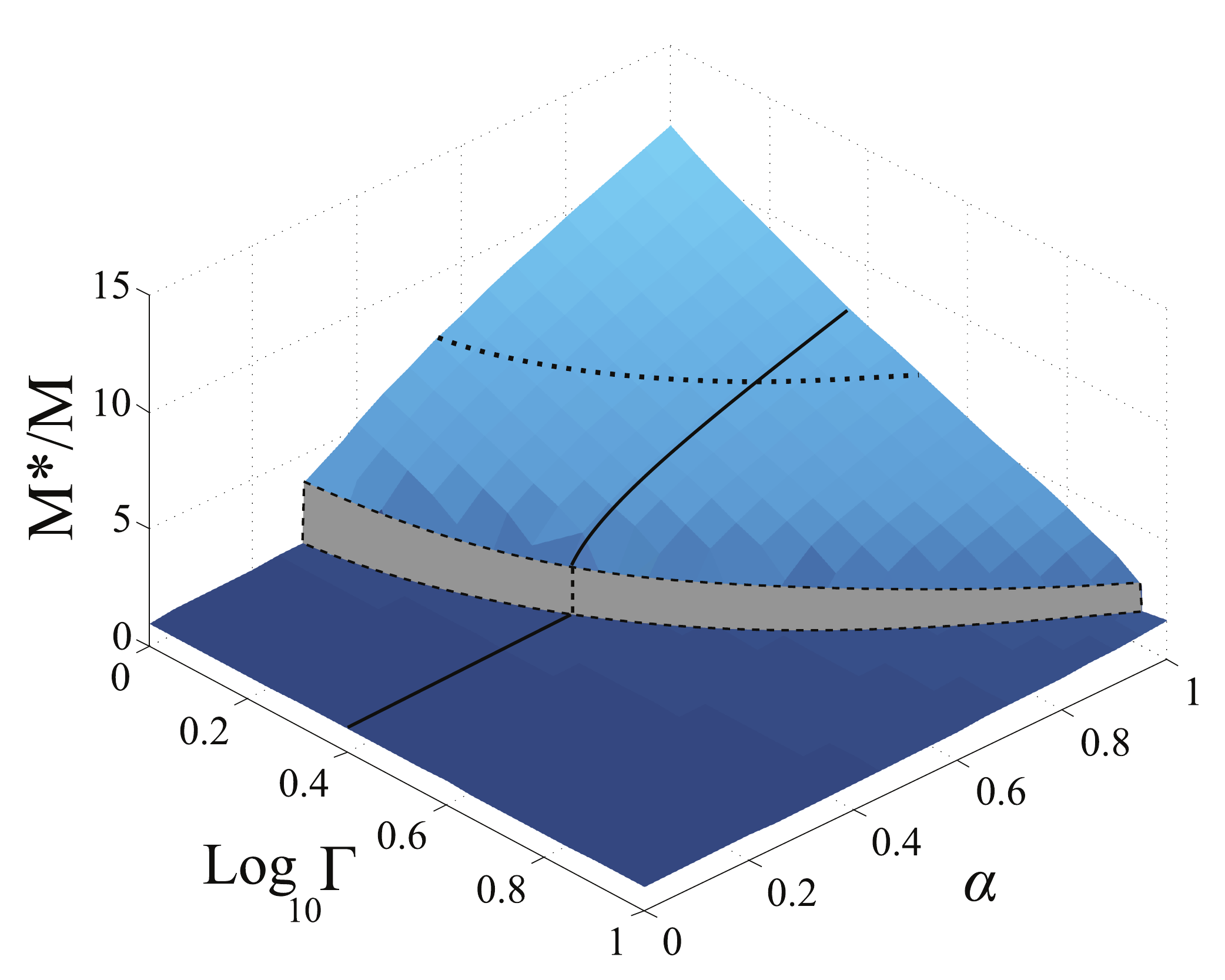}
\caption{\label{fig:sup2} Effective mass as a function of the coupling constant $\alpha$ and ${\text{Log}}_{10}\Gamma$ with $\Gamma$ the distance to the critical point of the Landau-Brazovskii model. The dotted line shows the locus of mass discontinuities predicted by Eq. \ref{weak}.}
\end{figure}
The location of self-trapping predicted by Eq. \ref{weak} agrees reasonably with the numerical results.

\subsection{Nonlinear Solutions}
{\it{Scaling}}.| The renormalized energy functional for a modulation pattern $A(r)$ for $\Delta\lesssim 0$ is given by:
\begin{align}
\UU\equiv &\int \dd^{D}r\bigg\{4q_0^2\bigg[\frac{\dd A(r)}{\dd r}\bigg]^2+\DDelta A(r)^2+\uu A(r)^4+wA(r)^6\nonumber\\
&+V(r)A(r)e^{i\qq_0.\rr}\bigg\},
\end{align}
where $D$ is the spatial dimensionality. The dimensions of the quantities are
\begin{subequations}
\begin{equation}
[A_0]=E^{1/2}L^{(2-D/2)},
\end{equation}
\begin{equation}
[q_0]=L^{-1},
\end{equation}
\begin{equation}
[\DDelta]=L^{-4},
\end{equation}
\begin{equation}
[\uu]=E^{-1}L^{D-8},
\end{equation}
\begin{equation}
[w]=E^{-2}L^{2D-12},
\end{equation}
\begin{equation}
[V_0]=\left[-\int \dd^D r V(r)e^{i\qq_0.\rr}\right]=E^{1/2}L^{D/2-2},
\end{equation}
\end{subequations}
(where $[E]$ denotes the dimension of energy). We will specialize to the transition point where $\uu^2=4w\DDelta$; there is only a single length scale left in the energy expression (apart from the impurity term) namely the correlation length $\xi\equiv2q_0/\sqrt{\DDelta}$. For modulation patterns that depend on position as $\tilde{r}=r/\xi$ the energy expression takes the form
\begin{align}
\UU=&\;\xi^D\int\dd^D \tilde{r}\bigg\{\frac{4q_0^2}{\xi^2}\bigg[\frac{\dd A}{\dd \tilde{r}}\bigg]^2+\DDelta A^2+\uu A^4+wA^6\nonumber\\
&+A\,V(\tilde{r}\xi)e^{i\qq_0.\tilde{\rr}\xi}\bigg\}.
\end{align}
Next, define a dimensionless modulation amplitude ${A}=(\DDelta/|\uu|)^{1/2} \tilde{A}$ (for $\uu<0$). The energy takes the form :
\begin{align}
\UU=&\;\frac{\DDelta^2}{|\uu|}\xi^D\int\dd^D \tilde{r}\bigg\{\bigg[\frac{\dd \Aa}{\dd \tilde{r}}\bigg]^2+\Aa^2-\Aa^4+\frac{1}{4}\Aa^6\nonumber\\
&+{\frac{|\uu|^{1/2}}{\DDelta^{3/2}}}\Aa\,V(\tilde{r}\xi)e^{i\qq_0.\tilde{\rr}\xi}\bigg\}.
\end{align}
where we used the fact that $\uu^2=4w\DDelta$ at the critical point.
For a potential with a range short compared to $\xi$, the amplitude can be expanded in the last term in a power series around the origin as $\Aa(\tilde{r})\simeq \Aa(0)+\frac{1}{2}{\tilde
{A}}^{\prime\prime}(0)\tilde{r}^2+...$ , in terms of which
\begin{align}
\UU=&\frac{\DDelta^2}{|\uu|}\xi^D\int\dd^D \tilde{r}\bigg\{\bigg[\frac{\dd \Aa}{\dd \tilde{r}}\bigg]^2+\Aa^2-\Aa^4+\frac{1}{4}\Aa^6\bigg\}\nonumber\\
&-\frac{\DDelta^{1/2}}{|\uu|^{1/2}}\bigg(\Aa(0)V_0+{\tilde{A}}''(0)\frac{V_2}{2\xi^2}+..\bigg),
\end{align}
where $V_0=-\int \dd^Dr V(r)e^{i\qq_0.\rr}$ and $V_2=-\int \dd^Dr\, r^2\, V(r)e^{i\qq_0.\rr}$ are even moments of the impurity potential. If $V(r)$ resembles a Gaussian with a width $a$, then $V_2$ is reduced in magnitude with respect to $V_0$ by a factor of order $(a/\xi)^2$ while the contribution of the $2m^{\text{th}}$ moment is reduced by a corresponding power $(a/\xi)^{2m}$. 

\subsection{Soliton Solutions}
The Euler-Lagrange equation of the quasi-classical energy is
\begin{equation}
\nabla^2 A-\frac{1}{2}\frac{\dd{\mathcal V}(A)}{\dd A}=0\,,
\end{equation}
where we temporarily drop tilde signs and where ${\mathcal V}(A)=A^2-A^4+\frac{1}{4}A^6$. In the absence of the impurity potential, the soliton corresponds to the solution $A(r/\xi)$ of this non-linear second order differential equation that interpolates between the two degenerate minima at $A(0)=\sqrt{2}$ and at $A(\infty)=0$. The associated energy is ${\UU}_0=\frac{\DDelta^2}{|\uu|}\xi^Dg_0$ where $g_0$ is a purely numerical factor. If the impurity potential at the origin is included then $\tilde{A}(0)$ no longer equals $\sqrt{2}$. The energy of the soliton has the general form ${\UU}(A(0))=\frac{\DDelta^2}{|\uu|}\xi^Dg(\tilde{A}(0))$ where $g(x)$ is a dimensionless scale function with $g(\sqrt{2})=g_0$. To learn about the form of $g(x)$, we specialize to the case of $D=1$ where the Euler-Lagrange equation can be viewed as the equation of motion of a fictitious particle with ``location" $A$, ``time" $r$, ``mass" one and ``potential energy" $-\mathcal{V}(A)$. By applying the principle of energy conservation, the function $g(x)$ is easily seen to equal
\begin{equation}
g(x)=2\int_0^x\dd y\sqrt{y^2-y^4+(1/4)y^6},
\end{equation}
as shown in Fig. [3].
\begin{figure}[ht]{\label{g}}
\centering
\includegraphics[width=0.6\linewidth]{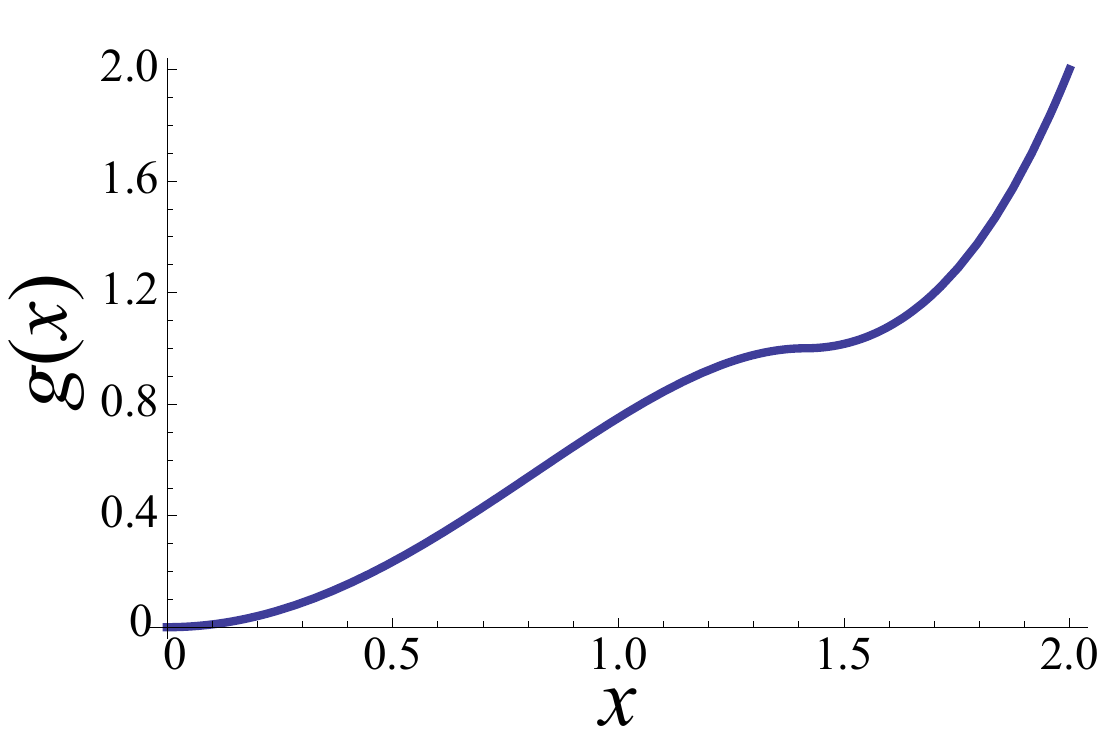}
\caption{\label{fig:sup2} Dimensionless energy $g(x)$ of a radial modulation profile as a function of the displacement at the origin.}
\end{figure}
The plot of $g(x)$ shows that this function is monotonic in $x$. For small $x$, $g(x)$ is proportional to $x^2$. With increasing $x$, the second derivative of $g(x)$ becomes negative. The slope decreases to zero at $x=\sqrt{2}$ where $g(x)$ has a cusp singularity. For larger $x$, the slope starts to increase again and the second derivative is positive once again. The intermediate region where $g''(x)$ is negative is unstable. For dimensions above one, this expression is no longer the exact scale function but the qualitative features remain the same.

\subsection{Derrick's Theorem}

Let ${A}({r})$ be a solution of the Euler-Lagrange equation (dropping  tilde signs), including the impurity potential. The energy of a modulation pattern that is stretched by a scale factor $\lambda$, so with ${A}(\lambda{r})$, is then given by
\begin{align}
{\UU}(\lambda)=\lambda^{(2-D)}\FF_1 +\lambda^{-D}\FF_2+\FF_3 +\lambda^2\FF_4 +... ,
\end{align}
where the coefficients
\begin{subequations}
\begin{equation}
\FF_1=\int\dd^D r\;\left(\frac{\dd A(r)}{\dd r}\right)^2,
\end{equation}
\begin{equation}
\FF_2=\int\dd^D r\;\mathcal V[A(r)],
\end{equation}
\begin{equation}
\FF_3=-V_0A(0),
\end{equation}
\begin{equation}
\FF_4=-\frac{1}{2}V_2A''(0),
\end{equation}
\end{subequations}
are all positive. The dots stand for higher-order even powers of $\lambda$, which we will drop in the following in which case $\;{\UU}\simeq\FF_1 +\FF_2+\FF_3 +\FF_4$ is the energy of the original soliton. As a function of the scale factor, ${\UU}(\lambda)$ has a single minimum that is stable. From the fact that ${A}(\lambda{r})$ is a solution of the Euler-Lagrange equations for $\lambda=1$, it follows that the minimum where $\dd\,{\UU}(\lambda)/\dd\lambda=0$ must be at $\lambda=1$ so
\begin{align}
(D-2)\FF_1 +D\FF_2-2\FF_4=0\,,
\end{align}
which can be viewed as a virial theorem. Note that in the absence of the impurity potential, the soliton solution is unstable for $D>1$, which is just Derrick's Theorem, but that the impurity potential suppresses the instability. Note also that the zeroth moment of the impurity potential does not enter in this expression. This condition fixes the curvature of the modulation profile at the origin to be:
\begin{align}
A''(0)=-\frac{(D-2)\FF_1 +D\FF_2}{V_2}<0\,.
\end{align}
Since the zeroth moment of the impurity potential is of the order of one at the transition point between linear response and the soliton solution, and since $\FF_{1,2}$ are positive numbers of the order of one, it follows that the curvature $|A''(0)|\sim(\xi/a)^2$ must be large.

\subsection{Gaussian Fluctuations}

In order to examine the stability of the soliton solution, we expand the action to quadratic order in fluctuations $\da$, so $\Aa\to \Aa+\da$. The free energy of the fluctuations reads:

\begin{equation}
\mathcal \delta F=2\beta\int\dd^3\rr\,\{\da(r)\hat{\mathcal K}\,\da(r)\},
\end{equation}
with $\hat{\mathcal K}=-\nabla^2+U(r)$ is the operator of a Schr\"{o}dringer equation with potential $U(r)=\frac{1}{2}\partial^2_{\Aa}\tilde{\mathcal V}[\Aa(r)]$. Stability requires all the eigenvalues of $\hat{\mathcal K}$ to be positive. When the field | at some distance from the origin | passes over the barrier of $\tilde{\mathcal V}(\Aa)$, the potential $U(r)$ becomes negative, because of the negative curvature of $\tilde{\mathcal V}(\Aa)$. The mean field solution is unstable if the ground state of the Schr\"{o}dinger equation is a negative energy bound state of this potential well. The ground state energy is greater than the minimum of $U(r)$ by $\frac{D}{2}\Omega$, where $\Omega$ is the frequency of small oscillations around the minimum of the potential $U(r_m)$, specifically, $\Omega=|\nabla^2 U|_{r=r_m}^{1/2}$, where $r_m$ is the minimum of $U$. In terms of the amplitude $\Aa$, the corresponding minimum is at $\Aa_m=\sqrt{4/5}$ (see the next figure). In order for there to be no bound states, we must demand $\frac{D}{2}\Omega$ to be larger than the barrier of $U(r)$. Also $U(r=0)=\frac{1}{2}\partial^2_{\Aa}\tilde{\mathcal V}(\Aa=\Aa_0)=4$ and $U(r\to\infty)=\frac{1}{2}\partial^2_{\Aa}\tilde{\mathcal V}(\Aa=0)=1$.

We have $\Omega=|\nabla^2 U|_{r=r_m}^{1/2}=\big[\frac{\partial^2 U}{\partial \Aa^2}\big(\frac{\dd \Aa}{\dd r}\big)^2\big]^{1/2}=\sqrt{24}\big|\frac{\dd \Aa}{\dd r}\big|$. We also know from Euler-Lagrange equation, that $\big[\frac{\dd}{\dd r}+\frac{2(D-1)}{r}\big]\big(\frac{\dd \Aa}{\dd r}\big)^2=\frac{\dd}{\dd r}\tilde{\mathcal V}(\Aa)$, hence $\big|\frac{\dd \Aa}{\dd r}\big|\leq|\tilde{\mathcal V}(\Aa)|^{1/2}=6\sqrt{5}/25$, and $\Omega\leq 2.74$, the equality is approached either in one dimension, or for large $r_m$ in higher dimensions where the effect of curvature is negligible. As shown in Fig. \ref{fig:U}, the minimum of the potential is $U_m=-1.4$. If $U_m+\frac{D}{2}\Omega>1$, the bound state does not form. This is valid for $D\geq2$ (for upper bound of $\Omega$).

\begin{figure}[ht]{\label{fig:U}}
\centering
\includegraphics[width=0.6\linewidth]{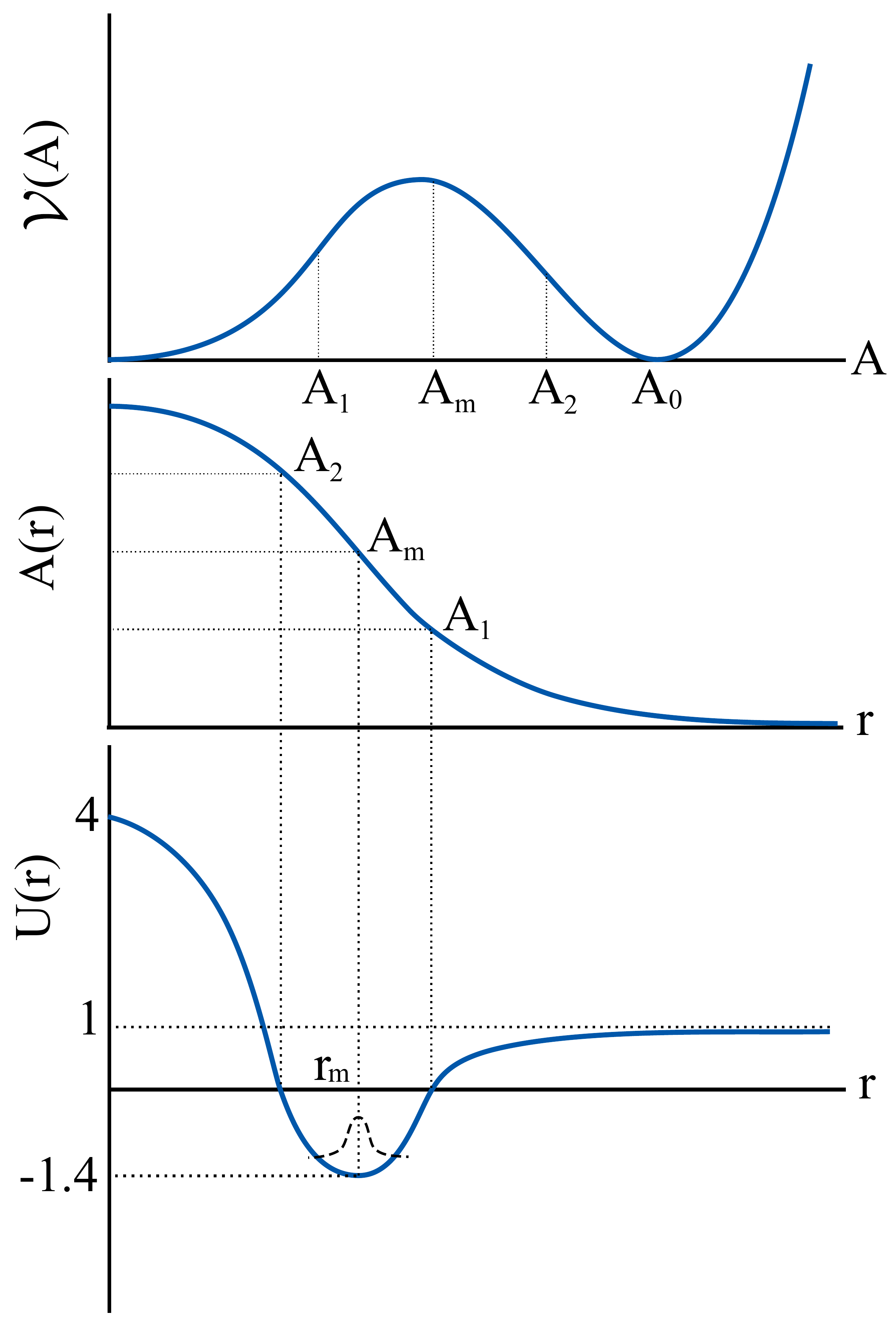}
\caption{\label{fig:U} (Tildes dropped from the labels) The top panel shows ${\mathcal V}(A)$ at the transition. Points $A_1$ and $A_2$, indicate where the curvature flips sign, whereas $A_m$ is where the curvature is minimum. The middle panel shows a schematic of the profile $A(r)$, taking all the values from $A_0$ down to $0$. The bottom one plots the potential $U(r)$ for GFs. The dashed Gaussian wave-packet shows the ground state of the fluctuations around the minimum of $U(r)$.}
\end{figure}


\begin{thebibliography}{100}
\bibitem{nonlinear} Odd-powered non-linear terms are excluded because of a selection rule imposed by the cavity. See Ref. \cite{Gopal2}.
\bibitem{ft1} Close to the critical point of the QLB transition, integrals in momentum space are dominated by the shell $q=1$.
\bibitem{ft2} In the original units, the classical structure factor is proportional to $k_BT$.
\bibitem{AP} A similar critical point is encountered in the theory of acoustic polarons \cite{Vlietinck}.
\end{thebibliography}
\end{document}